\documentclass[aps,amsmath,amssymb,preprint]{revtex4}
\usepackage{mathrsfs}
\usepackage{epsfig,bm,dcolumn}
\usepackage{graphicx}
\usepackage{color}
\usepackage{amsmath}

\begin{document}


\title{The induced interaction in a Fermi gas with a BEC-BCS crossover}

\author{Zeng-Qiang Yu, Kun Huang, Lan Yin}
\email{yinlan@pku.edu.cn}
\address{School of Physics, Peking University, Beijing 100871, China}

\date{\today}

\begin{abstract}
We study the effect of the induced interaction on the superfluid transition
temperature of a Fermi gas with a BEC-BCS crossover.  The Gorkov-Melik-Barkhudarov
theory about the induced interaction is extended from the BCS side to the entire
crossover, and the pairing fluctuation is treated in the approach by Nozi\`{e}res and Schmitt-Rink.
At unitarity, the induced interaction reduces the transition temperature by about twenty percent.
In the BCS limit, the transition temperature is reduced by a factor about 2.22, as found by Gorkov and Melik-Barkhudarov.
Our result shows that the effect of the induced interaction is important both on the BCS side and in the unitary region.
\end{abstract}



\maketitle
\section{Introduction}
One of the most important developments in experiments on ultra-cold atoms
is the observation of BEC-BCS crossover \cite{Jin03, Grimm03, Ketterle03,
Thomas04, Salomon04, Chin04, Ketterle05, Hulet05} which was originally
predicated for strongly-coupled superconductors \cite{Eagles69,
Leggett80}.  In experiments on ultra-cold atoms, the interaction
between atoms can be tuned by the technique of Feshbach resonance,
and the system can evolve smoothly from a Bardeen-Cooper-Schrieffer
(BCS) pairing state to a Bose-Einstein condensation (BEC) state of
diatomic molecules. In the BCS limit where the interaction is weakly
attractive, the atoms are paired into a BCS state below a critical
temperature, very similar to electrons in conventional
superconductors.  In the BEC limit where the interaction is weakly
repulsive, tightly-bound diatomic molecules are formed, and a
molecular BEC state appears below a critical temperature.  Near the
resonance, in the unitary region \cite{Ho04, Thomas05} where the
size of the scattering length is much larger than the inter-particle
spacing, the system is strongly correlated in both the normal and
superfluid states.

The BEC-BCS crossover can be qualitatively explained by a mean-field
BCS theory \cite{Leggett80}.  In this theory, the size of atom pairs
decreases as the system goes from the BCS side to the BEC side. In
the BEC limit, the pair size is so small that atom pairs become
diatomic molecules.   However, the mean-field theory predicts an
exponentially-divergent superfluid transition temperature in the BEC
limit \cite{Randeria93}, which is against the result from the theory
about an ideal Bose gas.  Nozi\`{e}res and Schmitt-Rink (NSR)
\cite{NSR85} first pointed out that the pairing fluctuation must be
taken into account to obtain the correct superfluid transition
temperature $T_c$ of the BEC-BCS crossover. The pairing fluctuation
is especially important in the BEC limit where nearly all atoms
become thermal molecules at $T_c$.  Following NSR's pioneer work,
many theoretical studies have focused on improving NSR's method and
extending their analysis to the broken symmetry state
\cite{Randeria93, Haussmann93, Randeria97, Chen98, Holland02,
Griffin02, Strinati04, Chen05, Hu06, Randeria08, Haussmann07}, which
was recently reviewed in Ref. \cite{Hu08, Levin08}.

However in the BCS limit, a different type of fluctuation is
important. Gorkov and Melik-Barkhudarov (GMB) \cite{GMB61} found
that there is a modification to the pairing interaction due to the
many-body medium, referred to as the induced interaction
\cite{Pethick00}. The induced interaction suppresses pairing
considerably and reduces the superfluid transition temperature $T_c$
by a factor about $2.22$ with respect to the mean-field $T_c$ in the
BCS limit. The fluctuation considered by GMB is in the particle-hole
channel, different from that in the particle-particle channel
considered by NSR.  In the BCS limit the NSR fluctuation is much
less important than the GMB fluctuation.  When the system moves from
the BCS side towards the BEC side, the GMB fluctuation becomes
weaker and the NSR fluctuation becomes stronger. In the BEC limit,
the NSR fluctuation is dominant. In an accurate description of the
BEC-BCS crossover, both GMB and NSR fluctuations should be treated
properly, which has not been addressed except in two recent
renormalization-group studies \cite{Floerchinger08, Stoof08}.

\begin{figure}
\includegraphics[width=8.5cm]{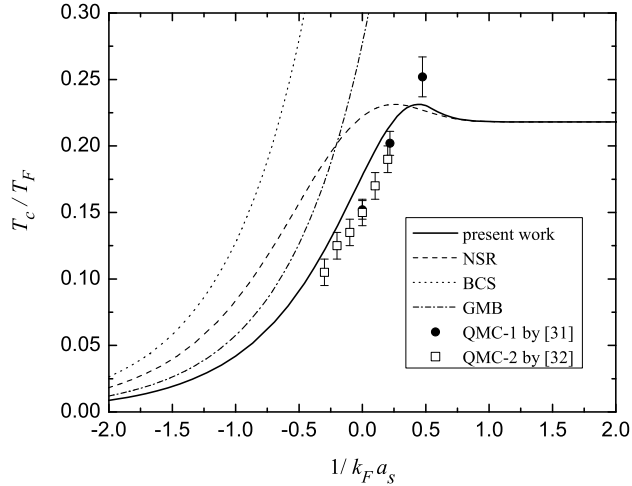}
\caption{Critical temperatures of BEC-BCS crossover.  The solid line
is our result after taking into account the induced interaction in
the NSR approach, the dashed line is the original NSR result, the
dotted line is the result from BCS mean-field theory, and the
dot-dashed line is GMB's result given by Eq. (\ref{GMB-Tc}).
These theoretical results are compared with the data
from QMC simulations \cite{Burovski08, Bulgac08} shown in
symbols.}\label{tc}
\end{figure}

In this work, we present our result about the induced interaction
in the whole BEC-BCS crossover.  First we extend the GMB
theory about the induced interaction from the BCS limit to the
strongly-interacting region.  Then we consider the induced
interaction in the NSR framework and compute $T_c$ for the entire
BEC-BCS crossover.  Our main result is shown in Fig. \ref{tc}.
Compared with the original NSR result, the superfluid transition temperature $T_c$  is
reduced considerably both on the BCS side and in the unitary region.  In the BCS limit,
we recover the GMB result.  At unitarity, the critical temperature $T_c$ is
found to be $T_c=0.178T_F$, close to the results from Quantum Monte Carlo (QMC)
simulations \cite{Burovski08, Bulgac08}, and about 20\% smaller than the NSR result.
Our result shows that induced interaction plays an important role in the unitary region
and on the BCS side.  Discussions and conclusions are given in the end.

\section{The induced interaction}
A Fermi gas with a wide Feshbach resonance can be described by a
single-channel model,
\begin{equation}
\mathcal{H}=-\sum_{\sigma} \frac{\hbar^2}{2m}\psi_{\sigma}^\dagger
\nabla^2\psi_{\sigma}+g\psi_{\uparrow}^{\dagger}
\psi_{\downarrow}^{\dagger}\psi_{\downarrow}\psi_{\uparrow},
\label{Hamiltonian}
\end{equation}
where the coupling constant is given by $g=4\pi \hbar^2 a_s/m$,
$a_s$ is the scattering length, and $\psi_\sigma$ is the field
operator for spin component $\sigma$. In this work, we consider only
the homogeneous spin-balanced case where the densities of
spin-$\uparrow$ and spin-$\downarrow$ atoms are the same.

\begin{figure}
\includegraphics[width=8.5cm]{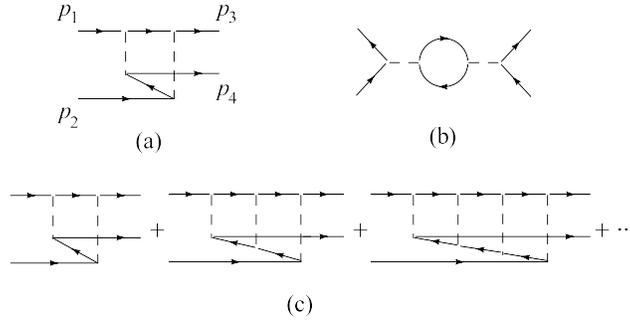}
\caption{Feynman diagrams in the particle-hole channel. (a) The induced
interaction considered by GMB. (b) An effective interaction between
atoms with the same spin component.  (c) The induced interaction in
the generalized GMB approximation. } \label{diagram2}
\end{figure}

In the BCS limit when the interaction is weakly attractive, Gorkov
and Melik-Barkhudarov (GMB) \cite{GMB61} showed that in the
particle-hole channel there is a correction to the pairing
interaction from the many-body background, given by the Feynman
diagram shown in Fig. \ref{diagram2}(a).  Other diagrams of the same
order are not as important. For example, the diagram in Fig.
\ref{diagram2}(b) corresponds to an effective interaction between
atoms with the same spin component, which is strongly suppressed at
low temperatures.

Beyond the BCS limit, when the interaction is strong, higher order
diagrams are important.  We generalize
the GMB approximation by considering all the diagrams of the same general type as
the GMB diagram shown in Fig. \ref{diagram2}(a) and summing all these
diagrams together as shown in Fig. \ref{diagram2}(c), and obtain the induced
interaction in the normal state given by
\begin{equation}
U_{\mathrm{ind}}( p_1, p_2; p_3, p_4)= -{g^2\, \chi(p_1-p_4) \over
1+g\chi(p_1-p_4)}, \label{induce}
\end{equation}
where $p_{i}=({\bf k}_{i}, \omega_{l_i})$ is a vector in the space
of wave-vector and frequency, and $\omega_{l}=(2l+1)\pi/(\hbar\beta)$ is
the Matsubara frequency of a fermion, $\beta=1/(k_BT)$.   The total
momentum and energy are conserved in the scattering,
$p_{1}+p_{2}=p_{3}+p_{4}$.  The function $\chi$ is taken as the
polarization function of a non-interacting Fermi gas with the same
chemical potential $\mu$, given by
\begin{eqnarray}
\chi(p')&=&\frac{1}{\hbar^2\beta V}\sum_{p} \mathcal{G}_0(p) \mathcal{G}_0(p+p') \nonumber \\
&=&\int {\mathrm{d}^3k \over (2\pi)^3} \frac{f_{{\bf k}}-f_{{\bf k}+{\bf
k}'}}{i\hbar\Omega_l+\epsilon_{{\bf k}}-\epsilon_{{\bf k}+{\bf k}'}},
\label{chi-ph}
\end{eqnarray}
where $p'=({\bf k}', \Omega_{l})$, $\Omega_{l}=2l\pi/(\hbar\beta)$ is
the Matsubara frequency of a boson, $V$ is the volume, $f_{\bf k}=1/[1+\exp(\beta \xi_{\bf k})]$
is the Fermi distribution function, $\xi_{\bf k}=\epsilon_{\bf
k}-\mu$, and $\epsilon_{\bf k}=\hbar^2k^2/2m$.  The
Green's function of a non-interaction Fermi gas $\mathcal{G}_0(p)$
is given by $\mathcal{G}_0(p)=\hbar/(i\hbar\omega_l-\xi_{\bf k})$.

Including the induced interaction, the effective
interaction between two atoms with different spin components is given by
\begin{eqnarray}
U_{\mathrm{tot}}( p_1, p_2; p_3, p_4)&=&g +U_{\mathrm{ind}}( p_1, p_2; p_3, p_4)\nonumber\\
&=&{g \over 1+g\chi(p_1-p_4)}. \label{induce}
\end{eqnarray}
Although the effective interaction is a function of transferred
momentum and frequency, at low temperatures only its s-wave part
plays an important role on pairing. As in GMB's work, we approximate
this s-wave component $g'$ by averaging the polarization function
\begin{equation}
g'={g \over 1+g\langle\chi\rangle}. \label{s-wave}
\end{equation}
When $\mu>0$, the average of the polarization function $\langle\chi\rangle$ is
obtained by setting the frequencies and total momentum to zero and
taking all the initial and final states of atoms from the Fermi
surface, i.e. ${\bf k}_1=-{\bf k}_2$, ${\bf k}_3=-{\bf k}_4$,
$k_1=k_2=k_3=k_4=k_F$, which yields
\begin{equation}
\langle\chi\rangle=\frac{m}{4\pi^2\hbar^2}
\int_{-1}^{1}\mathrm{d}\cos\theta\, \int_0^\infty \mathrm{d}k\,
\frac{k}{k'}f_{k}\ln\left|\frac{k'-2k}{k'+2k}\right|,
\label{avgchi-1}
\end{equation}
where $k'=|{\bf k}_1-{\bf k}_4|=k_F\sqrt{2(1+\cos\theta)}$, $k_F$ is
the Fermi wavevector, and $\theta$ is the angle between ${\bf k}_1$
and ${\bf k}_3$.  When the chemical potential $\mu$ turns negative on the BEC side,
the Fermi surface disappears, and the average $\langle\chi\rangle$ is taken
at zero frequency and in the limit that all the momentum go to zero,
$k_1=k_2=k_3=k_4\rightarrow0$, same as $k'\rightarrow0$ limit
of Eq. (\ref{avgchi-1}),
\begin{equation}
\langle\chi\rangle=-\frac{m}{2\pi^{2}\hbar^2}\int_{0}^{\infty}
\mathrm{d}kf_{k}.\label{avgchi-2}
\end{equation}
In both cases, the function $\langle\chi\rangle$ is always negative
and monotonically decreasing with the increase in the chemical
potential $\mu$.

In the BCS limit, the critical temperature $T_c$ is much less than
the Fermi temperature $T_F$.  Near $T_c$, one obtains
$$\langle\chi\rangle \approx -{\ln(4e)\over 3} \mathcal{N}(\epsilon_F),$$
where $\mathcal{N}(\epsilon_F)=mk_F/(2\pi^2\hbar^2)$ is the density
of states for one spin species at Fermi energy. The effective s-wave
interaction $g'$ is approximately given by
\begin{equation}
\frac{1}{g'} \approx \frac{1}{g}-{\ln(4e) \over
3}\mathcal{N}(\epsilon_F). \label{eff-int-weak2}
\end{equation}
With the effective pairing interaction $g'$, the GMB result of $T_c$
can be obtained,
\begin{equation}
T_c^{(\mathrm{GMB})} \approx
\left(\frac{2}{e}\right)^{7/3}\frac{\gamma}{\pi}T_F
e^{\pi/2k_Fa_s}\approx 0.28T_Fe^{\pi/2k_Fa_s}, \label{GMB-Tc}
\end{equation}
where $\gamma=e^{c}$, $c$ is the Euler constant.  The GMB result
$T_c^{(\mathrm{GMB})}$ is smaller by a factor of  $(4e)^{1/3}\approx
2.22$ than the mean-field $T_c$.

\section{The $T$-matrix and correction to density}
\begin{figure}
\includegraphics[width=8.5cm]{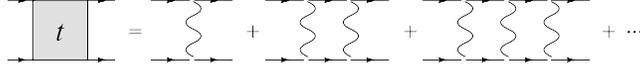}
\caption{Diagrams of the $T$-matrix.  The wiggled line represents
the effective s-wave interaction $g'$.} \label{diagram}
\end{figure}

To obtain the critical temperature for the whole BEC-BCS crossover,
we compute the $T$-matrix, as shown in Fig. \ref{diagram},
\begin{equation}
t(p')=\frac{g'}{1+g'\chi_{p}(p')}={1 \over
1/g+\langle\chi\rangle+\chi_{p}(p')}, \label{t-matrix}
\end{equation}
where the pair susceptibility $\chi_{p}(p')$  in particle-particle channel is given by
\begin{eqnarray}
\chi_{p}(p')&=&\frac{1}{\hbar^2\beta V}\sum_{p} \mathcal{G}_0(p) \mathcal{G}_0(p'-p)  \nonumber \\
&=&\int {\mathrm{d}^3k \over (2\pi)^3}\frac{f_{{\bf k}}+f_{{\bf
k}'-{\bf k}}-1}{i\hbar\Omega_l-\xi_{{\bf k}}-\xi_{{\bf k}'-{\bf k}}}.
\label{chi-pp}
\end{eqnarray}
Comparing with the conventional $T$-matrix approach, we have
replaced the coupling constant $g$ by the effective s-wave
interaction $g'$ due to the induced interaction.

According to Thouless criterion, the superfluid instability at $T_c$
is due to the divergence of $t(p'=0)$ which is equivalent to
\begin{equation}
\frac{m}{4\hbar^2\pi  a_s}+\langle\chi\rangle=\int {\mathrm{d}^3k
\over (2\pi)^3}\left(\frac{2f_{\bf k}-1}{2\xi_{\bf
k}}+\frac{1}{2\epsilon_{\bf k}}\right),\label{tc-eq}
\end{equation}
where the last term on the right-hand side is the counter term due
to vacuum renormalization.  Comparing the $T_c$-equation given by
Eq. (\ref{tc-eq}) with that in the BCS mean-field theory, the effect
of the induced interaction is equivalent to making the scattering
length larger.  In the BCS limit, the induced interaction leads to a
reduction of $T_c$ from the mean-field result, as given in Eq.
(\ref{GMB-Tc}).  In the BEC limit, the effect of the induced
interaction is negligible.

As Nozi{\`e}res and Schmitt-Rink pointed out \cite{NSR85}, pairing
fluctuations in the particle-particle channel are important
especially on the BEC side.  In the NSR theory, the total atom density
includes not only the fermion density in the mean-field approximation, but
also contributions from fluctuations of molecular fields.  In the
BEC limit, near $T_c$, the fluctuation contribution is dominant and
the critical temperature $T_c$ is given by the BEC temperature of an
ideal Bose gas.  One way to take into account the NSR effect is to
calculate the Hartree self-energy generated by the $T$-matrix and
its contribution to density  \cite{Chen05}. With our $T$-matrix
given by Eq. (\ref{t-matrix}) which includes the induced interaction,
the self energy is given by
\begin{equation}
    \Sigma(p)=\frac{1}{\hbar^2\beta V}\sum_{p'}t(p')\mathcal{G}_0(p'-p). \label{self-energy}
\end{equation}
To the first order, the Dyson's equation is given by $$
\mathcal{G}(p)=\mathcal{G}_{0}(p)+\mathcal{G}_{0}(p)\Sigma(p)\mathcal{G}_{0}(p),$$
and the particle density is given by
\begin{equation}
    n=\dfrac{2}{\hbar\beta V}\sum_{p}\mathcal{G}(p)e^{-i\omega_{l}0^{+}}=n_{f}+\Delta n  \label{density-eq},
\end{equation}
where the mean-field density is given by
    $$n_{f}=2\int f_{\bf k} {\mathrm{d}^3k \over (2\pi)^3},$$
and the fluctuation contribution $\Delta n$ is given by
\begin{equation}
\Delta n=\dfrac{2}{\hbar^3(\beta V)^{2}}\sum_{p'}\sum_{p}\mathcal{G}_{0}^{2}(p)\mathcal{G}_{0}(p'-p)t(p'). \label{density-flu1}
\end{equation}
If we omit the $\langle\chi\rangle$ term due to the induced
interaction in the $T$-matrix given by Eq. (\ref{t-matrix}), the
density equation given by Eq. (\ref{density-flu1}) is the same as
that in the NSR theory.

\section{The superfluid transition temperature}
 The superfluid transition temperature $T_c$ as a function of
the total atom density $n$ can be solved from the two coupled
equations (\ref{tc-eq}) and (\ref{density-eq}).  In the BCS limit,
since $n_f \gg \Delta n$, the GMB result about $T_{c}$ given by Eq.
(\ref{GMB-Tc}) can be recovered. In the BEC limit, at $T_c$, the
mean-field density is negligible, $\Delta n\gg n_f$, the $T$-matrix
is proportional to the propagator of noninteracting molecules, and
the density of total atoms is approximately given by density of
molecules.
Thus the transition temperature in the BEC limit is given by the
condensation temperature of an ideal Bose gas,
$T_{c}^{(\mathrm{BEC})}=0.218T_{F}$.

Our numeric result of the critical temperature for entire crossover
is shown in Fig. \ref{tc}. As expected, in the BCS limit, it agrees
with GMB theory; in the BEC limit, it recovers the condensation
temperature of ideal molecules. At unitarity, we obtain
$T_{c}=0.178T_{F}$, which is close to the QMC result
$T_{c}=0.15(1)T_{F}$ \cite{Burovski08, Bulgac08}. In comparison, the
results from other theoretical studies are $T_{c}=0.222T_{F}$ in the
original NSR theory, $T_{c}=0.160T_{F}$ in a full self-consistent
NSR treatment \cite{Haussmann07}, $T_{c}=0.26T_{F}$ in pseudogap
crossover theory \cite{Chen05}, $T_c=0.264T_F$ \cite{Floerchinger08}
and $T_{c}=0.13T_{F}$ \cite{Stoof08} in renormalization group
studies. Compared with the original NSR result, our critical
temperature is about 20\% lower, implying that the induced
interaction still plays an important role in the unitary region. The
chemical potential at $T_{c}$ in our results is
$\mu(T_{c})=0.598T_{F}$, higher than QMC results,
$\mu(T_{c})=0.493(14)$  \cite{Burovski08} and $\mu(T_{c})=0.43(1)$
\cite{Bulgac08}.  Our results can probably be improved by
self-consistently taking into account the self energy in the
computations of the induced interaction and $T$-matrix.  This issue
will be addressed in our further studies.

As reported in previous works \cite{NSR85, Randeria93, Holland02,
Strinati04, Haussmann07}, we also find that the critical temperature
reaches a maximum on the BEC side, as shown in Fig. \ref{tc}.  Compared
with the original NSR result, the position of this peak is
further away from the resonance due to the induced interaction.  Our results show that
the peak is located at $1/k_{F}a_{s}=0.437$, and $T_{c}^{\mathrm{peak}}=0.231T_{F}$,
close to the QMC estimation of the peak position
$1/k_{F}a_{s}\geq0.474(8)$ and
$T_{c}^{\mathrm{peak}}\geq0.252(15)T_{F}$ \cite{Burovski08}.

The effect of the fluctuation in the particle-hole channel on the
superfluid transition temperature was also studied in the
renormalization group approach for a two-channel model mostly in the
wide resonance case \cite{Floerchinger08} and for a single-channel
model \cite{Stoof08}. In the BCS limit, the superfluid transition
temperature was found in agreement with the GMB result in Ref.
\cite{Floerchinger08}, and smaller than the GMB result in Ref.
\cite{Stoof08} due to the simplification in the momentum dependence
of the interaction vertex. At unitarity, the superfluid transition
temperature was found to be $T_c=0.264T_F$ \cite{Floerchinger08} and
$T_{c}=0.13T_{F}$ \cite{Stoof08}, while we obtain $T_{c}=0.178T_{F}$
and the QMC result is $T_{c}=0.15(1)T_{F}$ \cite{Burovski08,
Bulgac08}.  On the BEC side when $k_Fa_s=0.5$, the superfluid
transition temperature was found to be $T_c \approx 0.25T_F$
\cite{Floerchinger08}, and our result shows $T_{c} \approx
0.22T_{F}$.  These quantitative differences may be resolved in
future studies with better theoretical treatments.

\section{Conclusion}
In conclusion, the effect of the induced interaction due to the
many-body medium is studied in a Fermi gas with the BEC-BCS
crossover.  The GMB theory is extended from the BCS limit to the
entire crossover. With the induced interaction considered, the
superfluid transition temperature $T_c$ is computed for the entire
crossover in the NSR framework.  The induced interaction reduces the
critical temperature $T_c$ considerably on the BCS side and in the
unitary region.  Our results of $T_{c}=0.178T_{F}$ at unitarity and
the $T_c$-peak location are in reasonable agreements with results
from quantum Monte Carlo simulations.   Our results show that the
effect of the induced interaction is important both in the unitary
region and on the BCS side.  We would like to thank T.-L. Ho for
helpful discussions. This work is supported by NSFC under Grant No.
10674007, and by Chinese MOST under grant number 2006CB921402.

\end{document}